\documentclass[10pt, letterpaper, twocolumn, twoside, conference]{IEEEtran}%a4paper, twocolumntwocolumn
\usepackage{dsfont}
\usepackage{cite}
\usepackage{graphicx}
\usepackage{amsmath}
\usepackage{amsfonts}
\usepackage{array}
\usepackage{times}
\usepackage{stfloats}
\usepackage{amssymb}
\usepackage{yhmath}
\usepackage{graphics}
\usepackage{textcomp}
\usepackage{amscd}
\usepackage{epsfig}
\usepackage{psfrag}
\usepackage{rotating}
\usepackage{amsmath}
\usepackage{url}
\usepackage{color}
\usepackage{float}
\usepackage{balance}

\newcommand{\expected}[2]{\ds\mathds{E}_{#2}\left[ #1 \right]}

\newcommand{\bs}{\boldsymbol}

\newcommand{\ds}{\displaystyle}

\newcommand{\pr}[1]{\mathrm{Pr}\left(#1\right)}

\begin{document}

\title{Strong Secrecy in Wireless Network Coding Systems with M-QAM Modulators}
\author{Arsenia Chorti$^\dag$, Mehdi M. Molu$^\flat$, David Karpuk$^\ddag$, Camilla Hollanti$^\ddag$, Alister Burr$^\flat$\\
\small{$^\dag$School of Computer Science and Electronic Engineering, Wivenhoe Park, Colchester, CO4 3SQ, UK}\\
\small{$^\flat$Department of Electronics, University of York, Heslington, York, YO10 5DD, UK} \\
\small{$^\ddag$Department of Mathematics and Systems Analysis, Aalto University, Helsinki, FI-00076, Finland }\\
achorti@essex.ac.uk, mhedi.molu@york.ac.uk, david.karpuk@aalto.fi, camilla.hollanti@aalto.fi, alister.burr@york.ac.uk}
\maketitle

\begin{abstract}
We investigate the possibility of developing physical layer network coding (PNC) schemes with embedded strong secrecy based on standard QAM modulators. The proposed scheme employs a triple binning approach at the QAM front-end of the wireless PNC encoders. A constructive example of a strong secrecy encoder is presented when a BPSK and an $8$-PAM modulator are employed at the wireless transmitters and generalized to arbitrary $M$-QAM modulators, assuming channel inversion is attainable at the first cycle of the transmission. Our preliminary investigations demonstrate the potential of using such techniques to increase the throughput while in parallel not compromise the confidentiality of the exchanged data.
\end{abstract}

\section{Introduction}
\begin{figure}[t]
\centering
\includegraphics[angle=0,width=0.3\textwidth]{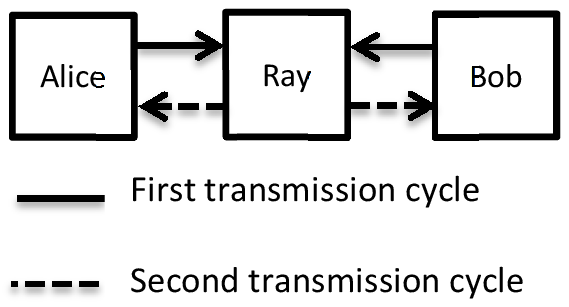}
\caption{Physical layer network coding (PNC) with two transmitter and one relay node.}
\label{fig:PNC}
\end{figure}

Recently, the ideas of network coding (NC) have been extended to the wireless physical medium; notably, in \cite{GastparIT}, \cite{Gastpar}, \cite{Popovski}, among others, the idea of harnessing interference through structured codes was explored in the framework of physical layer network coding (PNC). In such settings, the basic secrecy problem related to distributed communications is that two nodes want to communicate securely, but no direct link is available. Hence, they have to resort to  communicating via a potentially hostile relay, who should not gain any information on the transmitted secret messages. The generic system model with two independent sources and one relay is depicted in Fig. \ref{fig:PNC} and assumes that communication is executed in two cycles. In the first cycle, the nodes A, denoted as Alice, and B, denoted as Bob, transmit simultaneously codewords $\bs{x}_A$ and $\bs{x}_B$, respectively, to the relay node R, denoted as Ray. In the second cycle, Ray, transmits a function $f(\bs{x}_A+\bs{x}_B)$ of the received signals; Alice and Bob then retrieve each other's messages by canceling off their corresponding transmissions.

Depending on the transformation $f(\cdot)$ executed by Ray, one of the following relaying strategies can be employed \cite{GastparIT}:
\begin{itemize}
  \item \textit{Decode and forward}: Ray decodes the transmitted messages, at least partially and then transmits a re-encoded (aggregate) message. Such approaches have been shown to be interference limited.
  \item \textit{Compress and forward}: Ray is not required to decode the messages transmitted
but simply to describe its observation to the destinations. To this end, the observed signal is quantized and transmitted to the destinations.
  \item \textit{Amplify and forward}: Ray simply acts as a repeater. The main drawback of this approach is the aggregation of noise over the system. Such approaches have been shown to be noise limited.
  \item \textit{Compute and forward}: Recently, in \cite{GastparIT}, \cite{Gastpar} a novel strategy in which Ray decodes linear equations of the transmitted signals and forwards them to the destinations was introduced.
\end{itemize}

Earlier related work included the investigation of the secrecy capacity of interference channels in various settings \cite{He09}, \cite{PoorInterferer}, \cite{ChortiCISS12}. In the present contribution, we focus on the PNC setting and use a variation of the \textit{compress and forward} strategy when the channel state information (CSI) is available to Alice and Bob prior to their transmissions and $M$-QAM modulators are employed for the encoding of secret messages. Furthermore, we assume that the channel coefficients remain constant during each transmission slot.
During the first cycle Alice and Bob employ channel inversion strategies in their respective transmitters and broadcast $M$-QAM symbols. We propose the use of a triple binning of the QAM symbols at the source with the largest QAM constellation, hereafter assumed to be Bob without loss of generality. Each of Bob's $M$-QAM symbols is partitioned into:
\begin{enumerate}
  \item a bin of information bits that can be either secret bits (elements of messages that should be kept secret from Ray), or, common bits (elements of public messages intended to all receivers),
  \item a bin of bits used by Bob to enable the transmission of his secret bits,
  \item a bin of bits used by Bob to help Alice make a local decision regarding the transmission of her secret bits.
\end{enumerate}
The proposed technique is outlined with an example in section \ref{sec:Encoding}. Our goal is to investigate possible mapping schemes using QAM modulators that allow Ray to obtain estimates of linear combinations of the transmitted QAM symbols but not to retrieve any of the secret bits they carry, thus achieving strong secrecy (per QAM symbol instead of per transmission frame).

The paper is organized as follows. In section \ref{sec:system model} the system model is presented. In section \ref{sec:Encoding} a generic scheme that achieves strong secrecy is built, starting from a noiseless ideal scenario. Subsequently, this scheme is employed in section IV where a secure PNC architecture is proposed in a realistic channel setting. Finally in section V the conclusions of this contribution are drawn and future directions of the work are outlined.

\section{System model}\label{sec:system model}
Communication between Alice and Bob with the help of Ray takes place in transmission frames and is executed into two cycles. The first cycle from Alice and Bob to Ray spans $n$ transmission slots while the second cycle from Ray to Alice and Bob spans $m$ transmission slots, where $n, m$ are positive integers. In the first cycle Alice and Bob employ channel inversion at their respective transmitters while in the second cycle Ray employs a constant power policy. The transmission scheme is explained in further detail in subsections \ref{subsec: first cycle} and \ref{subsec: second cycle}.

\subsection{First transmission cycle}\label{subsec: first cycle}
In the first cycle Alice transmits to Ray a message $\bs{s}_A = [s_A(1), \ldots, s_A(q)] \in \mathcal{S}_A^{q}$, whose elements are uniformly drawn from a set of source symbols $ \mathcal{S}_A$. To this end, Alice employs an  encoding function $\varphi_A: \mathcal{S}_A^q \rightarrow \mathcal{X}_A^n$, with $\mathcal{X}_A$ a set of rectangular $M_A$-QAM symbols of size $m_A=\log_2(M_A)$ bits.
Each codeword is a sequence of $n$ QAM symbols denoted by $\bs{x}_A = \varphi_A(\bs{s}_A)$, $\bs{x}_A = \left[x_A(1), \ldots, x_A(n) \right]^T$, with average energy,
\begin{equation}\label{EqPowerConstraint}
\expected{|\bs{x}_A|^2}{}=\frac{1}{n}\ds\sum_{i = 1}^{n} \expected{|x_A(i)|^{2}}{} =\frac{1}{n}\ds\sum_{i = 1}^{n} \frac{M_A-1}{6} E_A(i),
\end{equation}
where $E_A(i)$ is the minimum distance of the QAM constellation during block $i$. Denoting by $h_A(i) $ the channel coefficient between Alice and Ray during block $i$, to implement channel inversion we pre-multiply the QAM symbols by $\frac{h_A^*(i)}{|h_A(i)|^2}$, so that,
\begin{equation}
E_A(i)=\frac{2}{|h_A(i)|^2}, i=1,\ldots, n.\label{eq:energy per bit Alice}
\end{equation}

At present, we assume that Alice is not power limited, i.e., it is always possible to transmit with energy per bit as described in (\ref{eq:energy per bit Alice}). We note that although channel inversion is impractical in Rayleigh environments, it can be employed whenever a line of sight (LOS) exists between either transmitter and Ray, i.e., whenever a Rician, a Nakagami-$m$ or other large scale fading channel model \cite{Hashemi} is applicable. In future work, we intend to investigate more general approaches regarding pre-equalization techniques at Alice and Bob, e.g., based on minimum mean square error (MMSE) pre-equalizers.

Similarly, Bob transmits to Ray a message $\bs{s}_B = [s_B(1), \ldots, s_B(p)] \in \mathcal{S}_B^{p}$, whose elements are uniformly drawn from a set of source symbols $ \mathcal{S}_B$. Bob employs an  encoding function $\varphi_B: \mathcal{S}_B^p \rightarrow \mathcal{X}_B^n$, with $\mathcal{X}_B$ a set of rectangular $M_B$-QAM symbols, each of length $m_B=\log_2(M_B)$ bits.
Each codeword at Bob is a sequence of $n$ $M_B$-QAM symbols denoted by $\bs{x}_B = \varphi_B(\bs{s}_B)$, $\bs{x}_B = \left[x_B(1), \ldots, x_B(n) \right]^T$, with average energy,
\begin{equation}\label{EqPowerConstraint}
\expected{|\bs{x}_B|^2}{}=\frac{1}{n}\ds\sum_{i = 1}^{n} \expected{|x_B(i)|^{2}}{} =\frac{1}{n}\ds\sum_{i = 1}^{n} \frac{M_B-1}{6} E_B(i),
\end{equation}
where $E_B(i)$ is the minimum distance of the QAM constellation during block $i$. Similarly, denoting by $h_B(i) $ the channel coefficient between Bob and Ray during block $i$, to implement channel inversion we pre-multiply the QAM symbols by $\frac{h_B^*(i)}{|h_B(i)|^2}$, so that,
\begin{equation}
E_B(i)=\frac{2}{|h_B(i)|^2}, i=1,\ldots, n.\label{eq:energy per bit Bob}
\end{equation}
Again, we assume that Bob is always able to transmit with energy per bit as described in (\ref{eq:energy per bit Bob}).

\par During transmission slot $i$ (corresponding to one channel coefficient), the signal received by Ray can be expressed as follows:
\begin{eqnarray}
y(i) &=& h_A(i) {x_A}(i) + h_B(i) x_B(i) +w(i),
\end{eqnarray}
where  $w(i)$ is a zero-mean circularly symmetric complex Gaussian random variable with variance $\sigma^2$. Thus, the received observation vector at Ray at the end of the first transmission cycle is given as:
\begin{equation}
\bs{y}=\bs{h}_A\bs{x}_A+\bs{h}_B\bs{x}_B+\bs{w}
\end{equation}
where $\bs{y}=[y(1),\dots,y(n)]^T$, $\bs{x}_A=[x_A(1),\ldots, x_A(n)]^T$, $\bs{x}_B=[x_B(1), \dots, x_B(n)]^T$, $\bs{h}_A=\mathrm{diag}(h_A(1), \ldots, h_A(n))$, $\bs{h}_B=\mathrm{diag}(h_B(1), \ldots, h_B(n))$, $\bs{w}=[w(1), \dots, w(n)]^T$.

At the Relay, the decoding functions $\phi: \mathcal{Y}^n \rightarrow \mathcal{S}^k$ and $\phi_A: \mathcal{Y}^n \rightarrow \mathcal{S}_A^q$, $\phi_B: \mathcal{Y}^n \rightarrow  \mathcal{S}_B^p$ are used to recover from the observations a function of the secret messages $\bs{s}_A,\bs{s}_B$ denoted by $\bs{s}=f(\bs{s}_A, \bs{s}_B)$ as well as the secret messages individually.  The error probabilities associated with the codes $(\varphi_A, \varphi_B,\phi)$, $(\varphi_A, \varphi_B,\phi_A)$ and $(\varphi_A, \varphi_B,\phi_B)$ are defined as:
\begin{eqnarray}
P_{e} &=& \pr{\phi(\bs{y}) \neq \bs{s}},\\
P_{e}^{(a)} &=& \pr{\phi_A(\bs{y}) \neq \bs{s}_A},\\
P_{e}^{(b)} &=& \pr{\phi_B(\bs{y}) \neq \bs{s}_B}.
\end{eqnarray}
%
%The level of ignorance of Ray with respect to the transmitted messages is measured by its equivocation rates $R_{e}^{(A)}$ and $R_{e}^{(B)}$ (weak secrecy conditions):
%%
%\begin{eqnarray}
%R_{e}^{(A)} &=& \frac{1}{n} H(\bs{S}_A|\bs{Y}) \text{ weak secrecy for Alice},\\
%R_{e}^{(B)} &=& \frac{1}{n} H(\bs{S}_B|\bs{Y}) \text{ weak secrecy for Bob},
%\end{eqnarray}
Strong strong secrecy can be achieved with respect to Ray if $\lim_{n\rightarrow \infty}I(\bs{S}_A; \bs{Y})=0, \lim_{n\rightarrow \infty}I(\bs{S}_B; \bs{Y})=0$. The secret messages can be transmitted by at rates $R_s^{(A)}$ and $R_s^{(B)}$, with
\begin{eqnarray}
R_s^{(A)} &\leq & H(\bs{S}_A|\bs{Y}),\\
R_s^{(B)} &\leq & H(\bs{S}_B|\bs{Y}).
\end{eqnarray}

\begin{table*}[t]
\begin{tabular}{ccc}
\includegraphics[angle=0,width=0.25\textwidth]{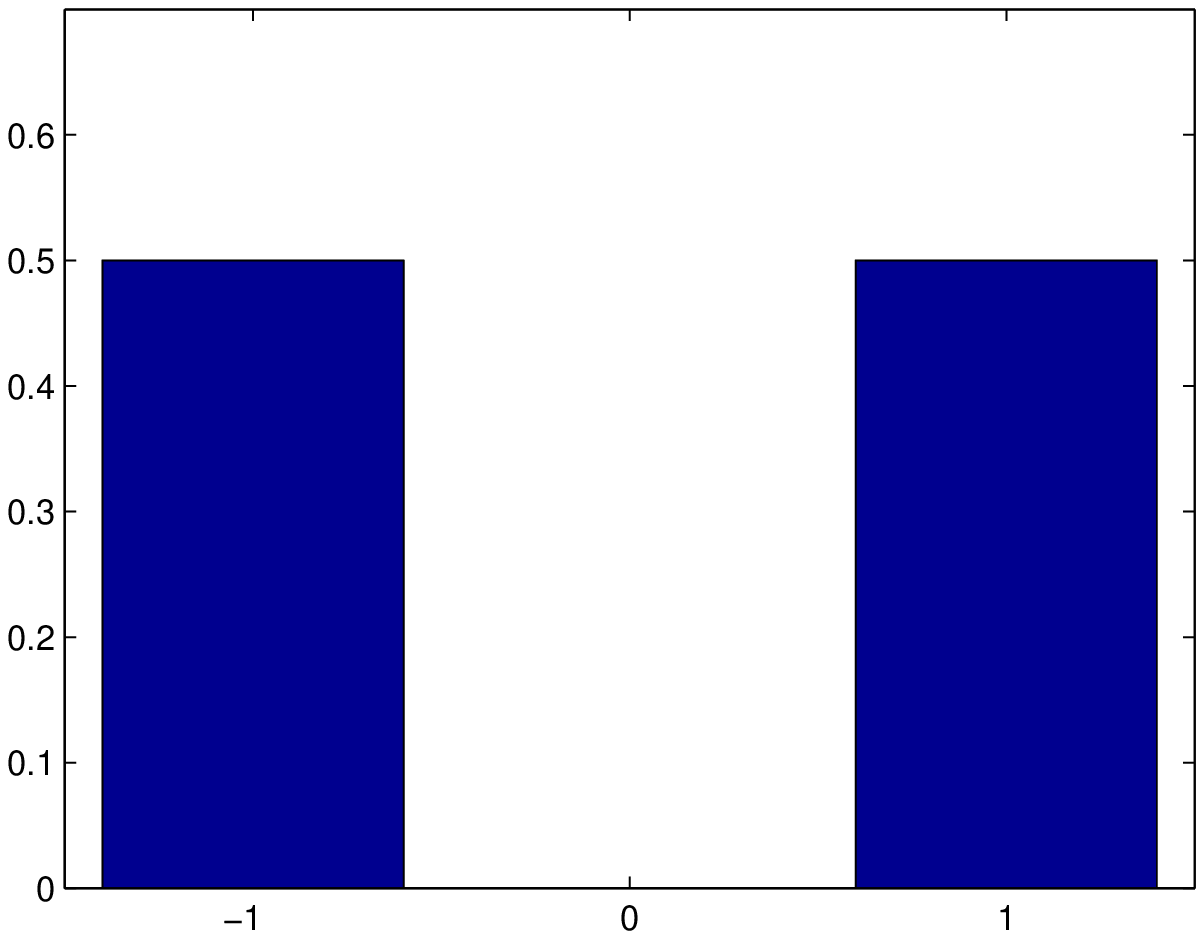}\label{fig:xa}
&
\includegraphics[angle=0,width=0.25\textwidth]{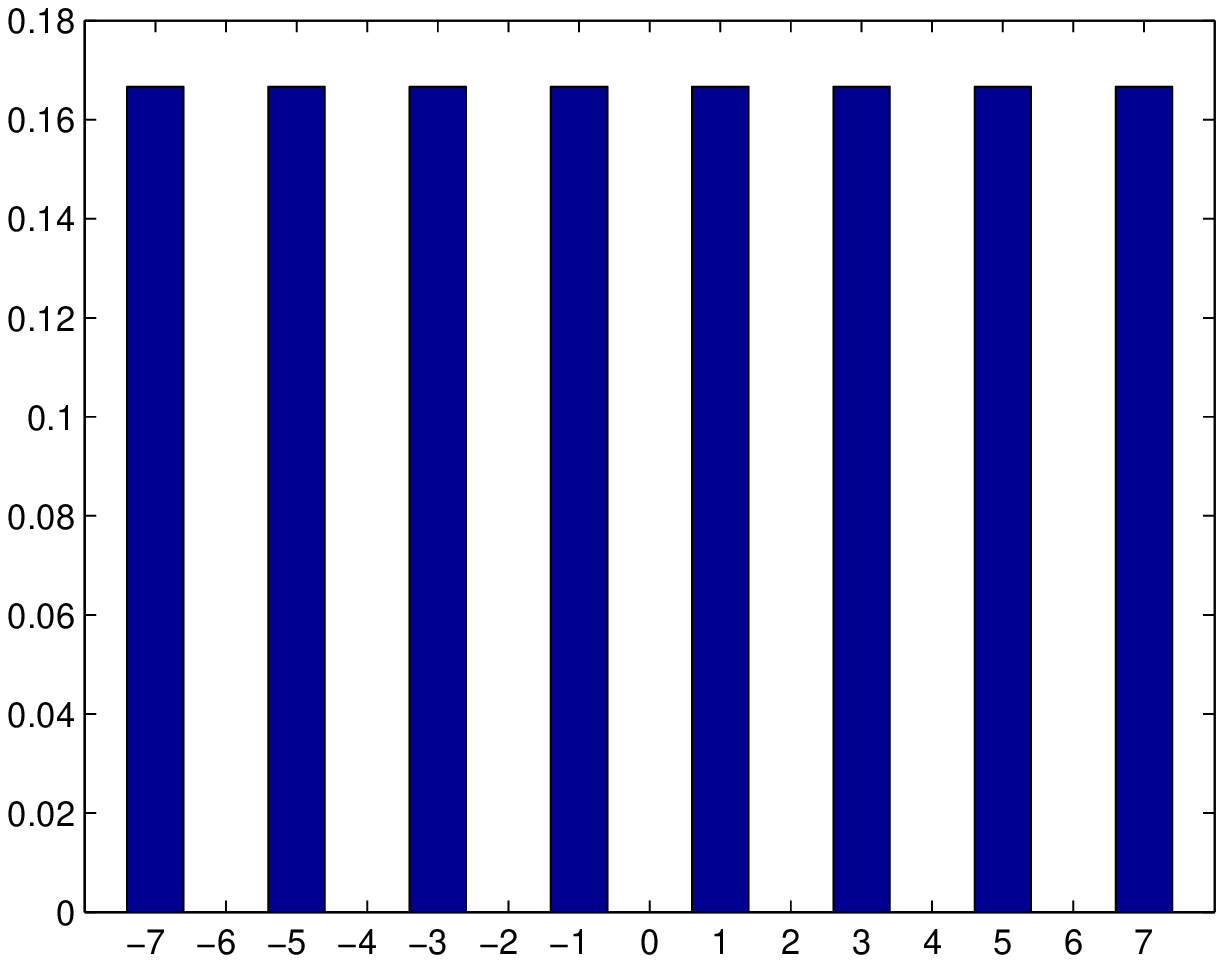}\label{fig:xb}
&
\includegraphics[angle=0,width=0.25\textwidth]{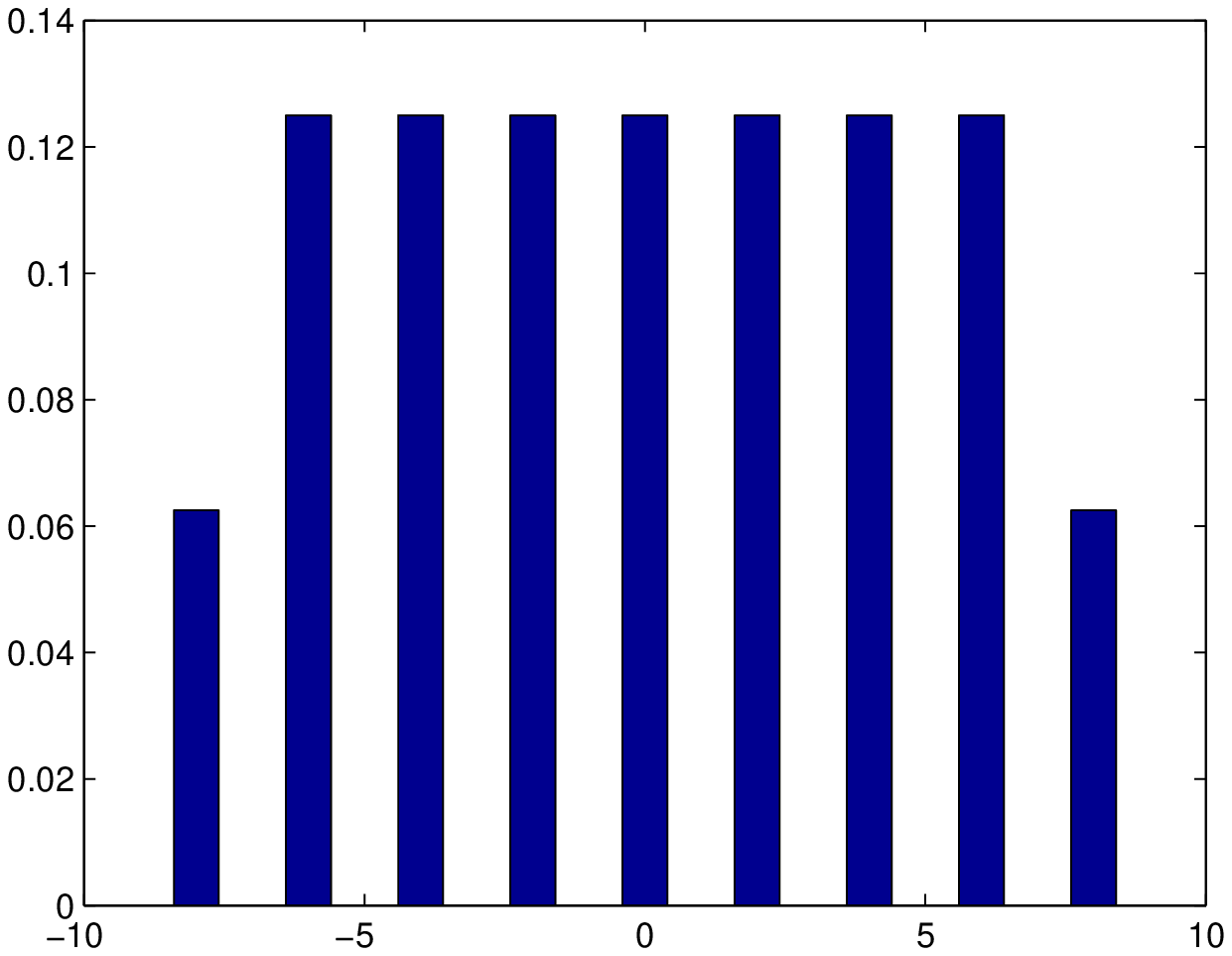}\label{fig:xa+xb}\\
(a) pmf of $x_A$ (BPSK) & (b) pmf of $x_B$ ($8$-PAM) & (c) pmf of $y=x_A+x_B$\\
& Fig. 2: pmfs of the transmitted and observed symbols &
\end{tabular}
\end{table*}
\setcounter{figure}{2}
\subsection{Second transmission cycle} \label{subsec: second cycle}
In the second cycle of the communication, Ray transmits PNC symbols $\bs{x}=[x(1), \dots, x(t)]^T$ to Alice and Bob. Towards this end, Ray employs an encoding function $\theta: \mathcal{S}^k \rightarrow \mathcal{X}^t$, with $\mathcal{X}$ a set of rectangular $M_R$-QAM symbols each of size $m_R=\max(m_A, m_B)$. Each codeword is a sequence of $t$ QAM symbols denoted by $\bs{x} = \theta(\bs{x})$, $\bs{x} = \left[x(1), \ldots, x(t) \right]^T$, with average energy,
\begin{equation}\label{EqPowerConstraint}
\expected{|\bs{x}|^2}{}=\frac{1}{n}\ds\sum_{i = 1}^{n} \expected{|x(i)|^{2}}{} =\frac{M_R-1}{6}E_R.
\end{equation}
Here for simplicity we set $E_R=2$, although a double waterfilling algorithm could be designed to optimize the power allocation across both communication links from Ray to Alice and from Ray to Bob. At present this is left as future work.

During the second part of the transmission, the channel coefficient between Alice and Ray in time slot $i$ is denoted by $\tilde{h}_{A}(i)$ while the channel coefficient between Bob and Ray is denoted by $\tilde{h}_{B}(i)$. In analogy to the first part of the transmission, the received observation vectors at Alice and Bob at the end of the second transmission cycle are given as:
\begin{eqnarray}
\bs{z}_A=\bs{\tilde{h}}_A\bs{x}+\bs{w}_A,\\
\bs{z}_B=\bs{\tilde{h}}_B\bs{x}+\bs{w}_B,
\end{eqnarray}
where $\bs{z}_A=[z_A(1),\dots,z_A(t)]^T$, $\bs{z}_B=[z_B(1),\dots,z_B(t)]^T$, $\bs{x}=[x(1),\ldots, x(t)]^T$,  $\bs{\tilde{h}}_A=\mathrm{diag}(\tilde{h}_A(1), \ldots, \tilde{h}_A(t))$, $\bs{\tilde{h}}_B=\mathrm{diag}(\tilde{h}_B(1), \ldots, \tilde{h}_B(t))$, $\bs{w}_A=[w_A(1), \dots, w_A(t)]^T$, $\bs{w}_B=[w_B(1), \dots, w_B(t)]^T$.

Furthermore, the decoding functions $\theta_A:  \mathcal{Z}_A^t \rightarrow \mathcal{S}_B^q$ and  $\theta_B:  \mathcal{Z}_B^t \rightarrow \mathcal{S}_A^p$ are used by Alice and Bob respectively to decode the individual messages transmitted from Bob and Alice based on the PNC observations.
The error probabilities associated with the codes $(\varphi_A, \phi, \theta, \theta_A)$ and $(\varphi_B,\phi, \theta, \theta_B)$ during a complete transmission frame are defined as
\begin{eqnarray}
P_{e}^{(A)} &=& \pr{\theta_A(\bs{z}_A) \neq \bs{s}_B| \bs{s}_A},\\
P_{e}^{(B)} &=& \pr{\theta_B(\bs{z}_B) \neq \bs{s}_A| \bs{s}_B}.
\end{eqnarray}

\par In the following, we focus on information theoretic strong secrecy; we are interested in building encoders at rates $R_s^{(A)}$ and $R_s^{(B)}$ (at Alice and Bob, respectively), such that for small $\epsilon_1, \epsilon_2, \epsilon_3 > 0$ the following conditions hold:
\begin{eqnarray}
\label{EqProbErrorCondition}
 P_{e} & \leqslant & \epsilon_1, \\
 \label{EqProbErrorConditionA}
 P_{e}^{(A)} & \leqslant & \epsilon_2, \\
 \label{EqProbErrorConditionB}
 P_{e}^{(B)} & \leqslant & \epsilon_3, \\
 R_s^{(A)} &\leq & H(\bs{S}_A|\bs{Y}) \label{eq: strong secrecy Alice},\\
R_s^{(B)} &\leq & H(\bs{S}_B|\bs{Y}) \label{eq:strong secrecy Bob}.
\end{eqnarray}
To demonstrate how strong secrecy can be achieved, we start by examining the noiseless scenario and concentrate on the design of mapping functions that satisfy (\ref{eq: strong secrecy Alice}) and (\ref{eq:strong secrecy Bob}) with equality.

\section{Achieving Strong Secrecy with Triple Binning QAM Modulators} \label{sec:Encoding}
In subsections \ref{subsec: basic} and \ref{subsec:arbitrary PAM} we present a constructive approach to develop the required encoders. We start with a basic example in which asymptotically $1$ bit can be transmitted with strong secrecy assuming that Alice employs a BPSK modulator and Bob an $8$-PAM modulator, in the absence of noise sources and ideal channel conditions. The proposed technique is generalized to arbitrary $M_A$-QAM and $M_B$-QAM modulators in subsection \ref{subsec:arbitrary PAM}.
\subsection{Basic scenario}\label{subsec: basic}
 At present we neglect all noise sources and set all chancel gains equal to unity. In our approach we employ a BPSK modulator at Alice transmitting symbols $x_A$, and an $8$-PAM modulator at Bob transmitting symbols $x_B$.
The constellations of the BPSK and the $8$-PAM are assumed as shown in Table \ref{ta:M-PAMs}.
\begin{table}[t]
\renewcommand{\arraystretch}{1.3}
  \centering
  \caption{$M$-PAM constellations}\label{ta:M-PAMs}
\begin{tabular}{l|l}
 \hline
 $M$ & Mapping to constellation points \\
 \hline
 $2$ &  $1$, $0$\\
 %$4$ & $11$, $10$, $00$, $01$ \\
 $8$ & $111$, $110$, $100$, $101$, $001$, $000$, $010$, $011$ \\
 %$16$ & $1111$, $1110$, $1100$, $1101$, $1001$, $1000$, $1010$, $1011$, $0011$, $0010$, $0000$, $0001$, $0101$, $0100$, $0110$, $0111$ \\
\hline
 \end{tabular}
\end{table}
Based on our system model, during each transmission block Ray simply observes the sum of a BPSK symbol and an $8$-PAM symbol, i.e.,
\begin{equation}
y=x_A+x_B.
\end{equation}
The probability mass functions (pmf) of the BPSK, the $8$-PAM and of Ray's observation are depicted in Fig. 1(a), 1(b) and 1(c), respectively.
Regarding the decoding of $x_A$ and $x_B$ by Ray, with probability $0.0625$ it can be correctly decoded as $(\hat{x}_A, \hat{x}_B)=(-1, -7)$ when $y=-8$ and with probability $0.0625$ as $(\hat{x}_A, \hat{x}_B)=(1, 7)$ when $y=8$. In all other cases $x_A$ cannot be decoded.
\begin{table*}[t]\label{ta:Ray's observation}
\centering
\caption{Ray's observation, detection and generation of the PNC codeword}
    \begin{tabular}{|c||c|c|c|c|c|c|c|c|c|}
    \hline
    $\bs{y}$  & $-8$ & $-6$ & $-4$ & $-2$ & $0$  & $+2$ & $+4$ & $+6$ & $+8$ \\
     \hline
    $(\bs{x}_A,\bs{x}_B)$  & $(-1,-7)$ & $(-1,-5)$ & $(-1,-3)$ & $(-1,-1)$ & $(-1,+1)$  & $(-1,+3)$ & $(-1,+5)$ & $(-1,+7)$ &--- \\
    & --- & $(+1,-7)$ & $(+1,-5)$ & $(+1,-3)$ & $(+1,-1)$  & $(+1,+1)$ & $(+1,+3)$ & $(+1,+5)$ & $(+1,+7)$ \\
\hline
    $\bs{s}= f(\bs{y})$ & $-7$ & $-5$ & $-3$ & $-1$ & $+1$ & $+3$ & $+5$  & $+7$  & $-7$ \\
\hline
    \end{tabular}
\end{table*}
This effect is explained in Table \ref{ta:Ray's observation}, where also a ``re-wrapping'' approach for Ray's PNC encoding scheme $\bs{s}= f(\bs{y})$ is presented.

We assume that both Alice and Bob have separate queues of ``secret'' ($s$) and ``common'' ($c$) bits, denoted by $\mathcal{Q}_s^{(A)}$ and $\mathcal{Q}_c^{(A)}$ for Alice and $\mathcal{Q}_s^{(B)}$  and $\mathcal{Q}_c^{(B)}$  for Bob respectively. During each transmission interval Alice and Bob make local decisions to transmit either a bit from $\mathcal{Q}_s^{(A)}$ or $\mathcal{Q}_c^{(A)}$ and $\mathcal{Q}_s^{(B)}$  or $\mathcal{Q}_c^{(B)}$ respectively. Their actions do not necessarily coincide, i.e., Bob might decide to transmit a bit from $\mathcal{Q}_s^{(B)}$  while Alice decides to transmit a bit from  $\mathcal{Q}_c^{(A)}$ and vice versa.

 Furthermore, we note that as depicted in Fig. 1(c) secrecy cannot be achieved neither by Bob nor by Alice when Bob transmits either of his $8$-PAM edge constellation points, i.e., when Bob transmits x$11$, with x$\in \{0, 1\}$. In all other occasions \textit{both} Alice and Bob can transmit one secret bit each. Therefore, whenever Bob transmits x$11$, then x should be drawn from $\mathcal{Q}_c^{(B)}$ and also Alice should transmit a bit from $\mathcal{Q}_c^{(A)}$. In all other cases Bob and Alice can securely transmit one bit each from their respective queues $\mathcal{Q}_s^{(B)}$ and $\mathcal{Q}_s^{(A)}$. To exploit this effect, the central idea in our approach is to use Alice's BPSK transmission to potentially ``mask'' the first of Bob's bits while use the remaining two bits in Bob's QAM symbol for indexing.

As a result, Bob's QAM symbols are partitioned into three bins. The first of Bob's bits is mapped to a bin of information bits.
 The remaining two bits are split into two partitions $\mathcal{K}_{B}$ and $\mathcal{K}_{A}$,
 of sizes $K_B=|\mathcal{K}_B|$ and $K_A=|\mathcal{K}_{A}|$. The bits in partition $\mathcal{K}_B$ are used by Bob to index his transmission when a secret bit from $\mathcal{Q}_s^{(B)}$ or when a common bit from $\mathcal{Q}_c^{(B)}$ is sent over the wireless channel. In analogy the bits in partition $\mathcal{K}_A$ are used by Bob to notify Alice whether in the \textit{next} transmission slot she should be transmitting a bit from the queue $\mathcal{Q}_s^{(A)}$ or from the queue $\mathcal{Q}_c^{(A)}$; we assume that Alice always uses this information (when available) to make her local decision. Naturally, since one out of three bits is used to carry information,
   \begin{equation}
   K_A+K_B\leq2.
   \label{eq:coset sizes}
   \end{equation}
There are two options regarding the use of the indexing bits.

\textit{Option 1: $K_B=2, K_A=0$:} A first option is to use the last two bits of the $8$-PAM symbol to index only Bob's transmission, i.e., $K_B=2$ (which implies $K_A=0$), so that Bob transmits a common bit when his index is $11$ and a secret bit in all other cases, i.e., Bob can achieve a secrecy rate $R_s^{(B)}=\frac{6}{8}=\frac{3}{4}$. In this scheme there is no feedback to Alice who cannot transmit with any \textit{guaranteed} strong secrecy, i.e., $R_s^{(A)}=0$.

\textit{Option 2: $K_B=1, K_A=1$:} As a second option, if the second of Bob's bits is used as an index for Bob and the third of the bits is used as an index for Alice, i.e., $K_B=1$ and $K_A=1$, then Bob shall transmit a common bit whenever his index is $1$ and a secret bit when his index is $0$. This in turn means that Bob will transmit a bit from $\mathcal{Q}_c^{(B)}$ when either of the four edge constellation points are transmitted, i.e., when either x$11$ or x$10$ is transmitted and a bit from $\mathcal{Q}_s^{(B)}$ in all other cases, leading to a secrecy rate of $R_s^{(B)}=\frac{4}{8}=\frac{1}{2}$. In parallel, Alice is notified whether she should transmit a common bit in the next transmission interval when her indexing bit is $1$ or a secret bit when her indexing bit is $0$, leading to a secrecy rate of $R_s^{(A)}=\frac{4}{8}=\frac{1}{2}$.

\begin{figure}[t]
\centering
\includegraphics[angle=0,width=0.50\textwidth]{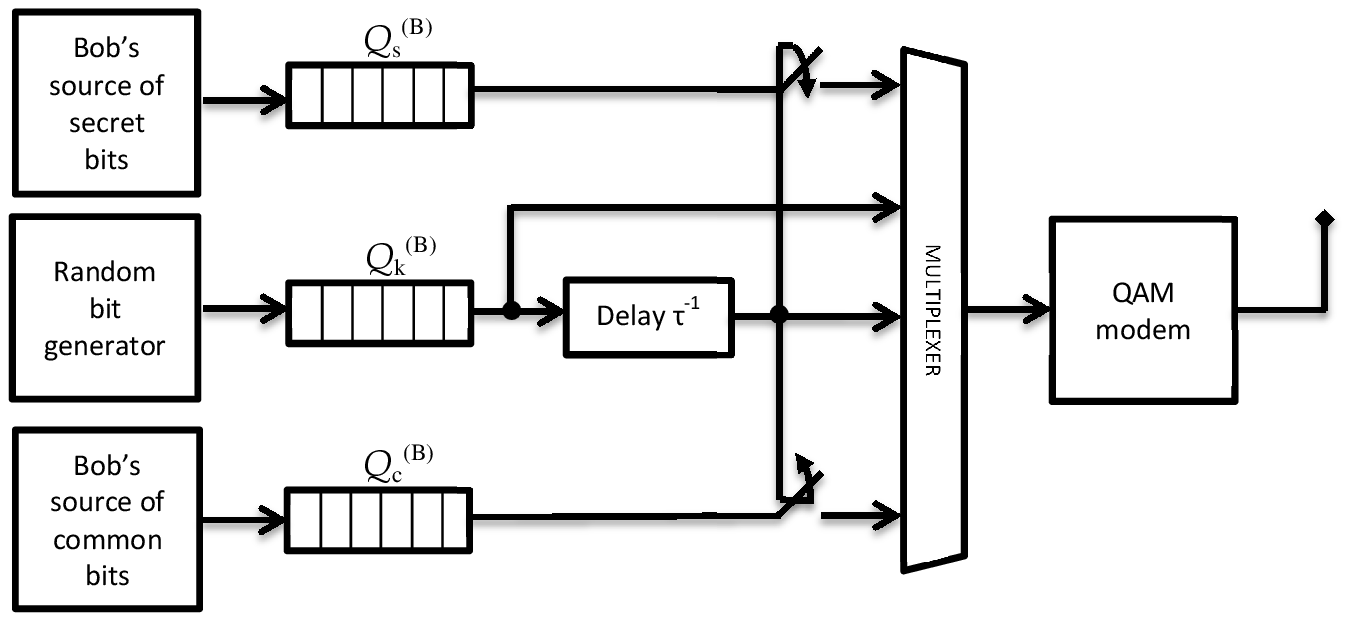}
\caption{Architecture of the basic scheme.}
\label{fig:basic scheme}
\end{figure}

Alice's index bit on slot $i$ is related to Bob's index bit on slot $i+1$; in particular, in the specific scheme these should coincide. A simple implementation of the approach is depicted in Fig. \ref{fig:basic scheme} . During an initialization slot a sequence of random bits is generated at Bob and stored in a queue $\mathcal{Q}_k^{(B)}$. These bits are used as index bits for Alice (bits in partition $\mathcal{K_A}$). A shifted version of this sequence is also used as index bits for Bob (bits in partition $\mathcal{K}_B$). Finally, the latter control a switch between Bob's queues $\mathcal{Q}_s^{(B)}$ and $\mathcal{Q}_c^{(B)}$. We note in passing that it can be shown that Bob's symbols's distribution remains uniform.

\begin{figure}[t]
\centering
\includegraphics[angle=0,width=0.50\textwidth]{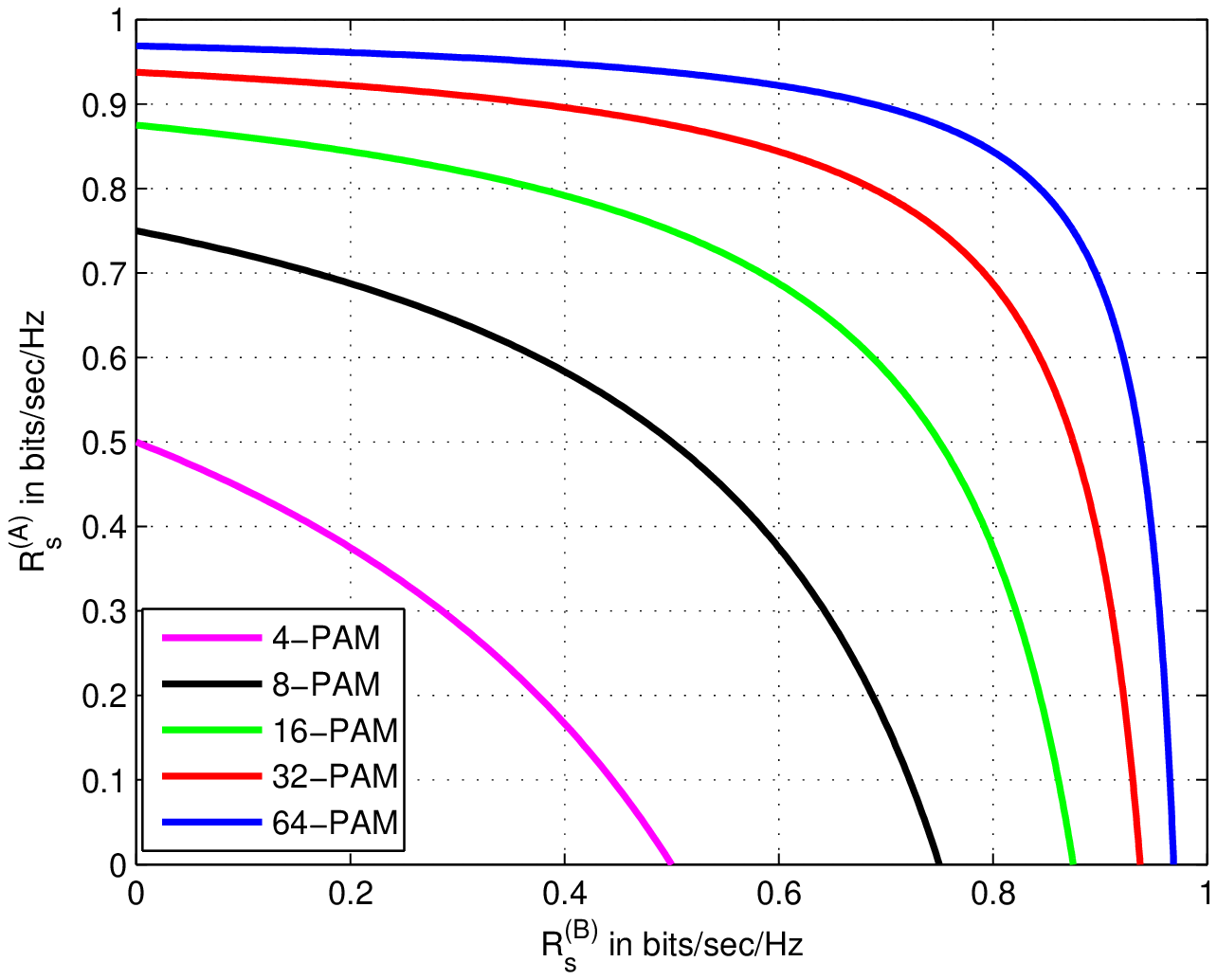}
\caption{Secrecy rates at Bob and Alice using a BPSK and an $M$-PAM modulators.}
\label{fig:Rs_PAM_BPSK}
\end{figure}

Using timesharing techniques we can let the sizes of the partitions $K_A$ and $K_B$ take any values between $[0, 2]$ as long as they satisfy (\ref{eq:coset sizes}). The secrecy rates that are achievable when employing an $8$-PAM modulator and a BPSK modulator can thus be expressed as:
\begin{eqnarray}
R_s^{(A)}&\leq&1-2^{-K_A},\\
R_s^{(B)}&\leq&1-2^{-K_B}.
\end{eqnarray}
In Fig. \ref{fig:Rs_PAM_BPSK} the achievable secrecy rates are depicted for a BPSK and an $M=4, 8, 16, 32, 64$-QAM modulators at Alice and Bob. As $M$ increases, we can asymptotically transmit $1$ bit/sec/Hz with strong secrecy.

\begin{figure}[t]
\centering
\includegraphics[angle=0,width=0.50\textwidth]{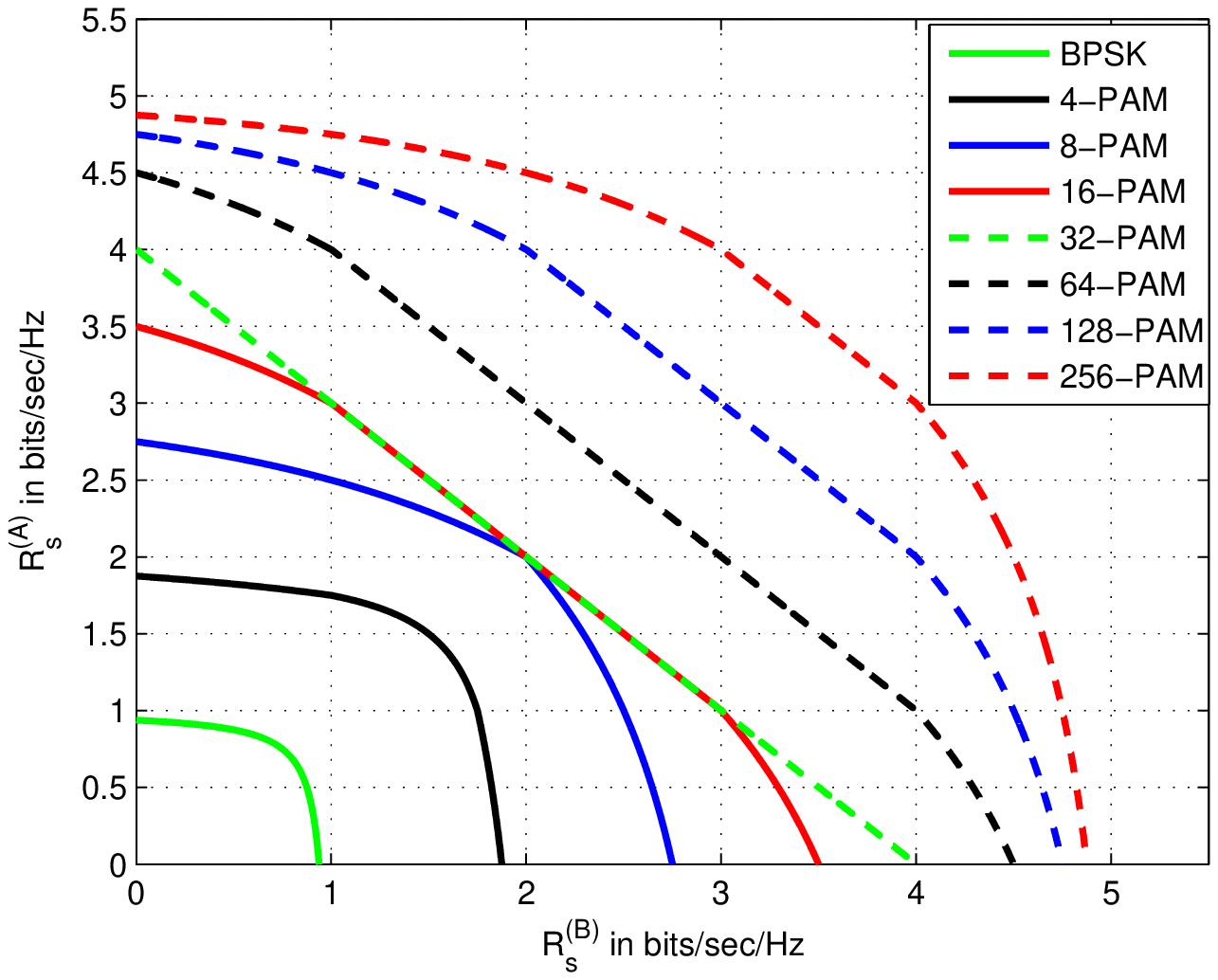}
\caption{Secrecy rates at Bob and Alice using an $M_A$-PAM and an $M_B$-PAM modulators with $M_B=64$.}
\label{fig:Generic rates}
\end{figure}
\begin{figure}[t]
\centering
\includegraphics[angle=0,width=0.50\textwidth]{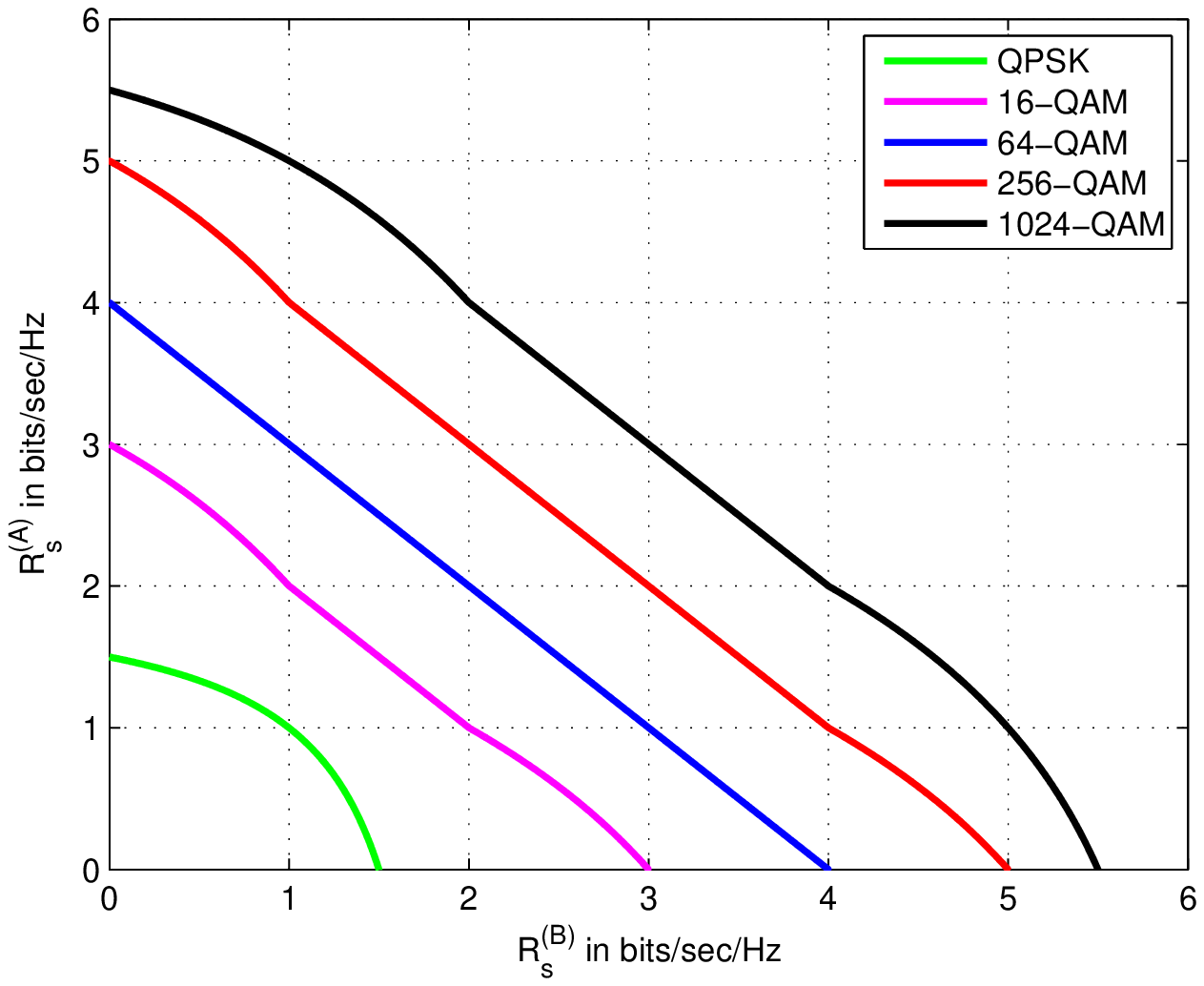}
\caption{Secrecy rates at Bob and Alice using an $M_A$-QAM and an $M_2$-QAM modulators with $M_B=64$.}
\label{fig:QAM rates}
\end{figure}

\subsection{Generalization to arbitrary PAM and QAM modulators}\label{subsec:arbitrary PAM}
We generalize the above methodology to propose triple binning encoders based on using an $M_A$-PAM modulator at Alice and an $M_B$-PAM modulator at Bob. We denote the lengths of the PAMs by $m_A=\log_2(M_A)$ and $m_B=\log_2(M_B)$ respectively and assume that $M_A\leq M_B$.
\par\noindent\textit{Proposition 1:} Using the proposed triple binning approach, the maximum achievable secrecy rates by Bob or Alice are upper bounded by $\tau=\min(m_A, m_B)$:
\begin{eqnarray}
R_s^{(A)}\leq \tau, \text{ and } R_s^{(B)}\leq \tau. \label{eq:Rs}
\end{eqnarray}
\par\noindent \textit{Sketch of proof:} The proof of (\ref{eq:Rs}) is straightforward as a result of the data processing theorem:
\begin{eqnarray}
R_s^{(A)}&=&H(\bs{S}_A|\bs{Y})\leq H(\bs{S}_A)\leq\tau, \\
R_s^{(B)}&=&H(\bs{S}_B|\bs{Y})\leq H(\bs{S}_B)\leq \tau.
\end{eqnarray}
\par\noindent\textit{Proposition 2:} The limit $R_s^{(A)}= \tau$ and $R_s^{(B)}= \tau$ can be achieved when the largest of the constellations becomes arbitrarily long, i.e., when $M_B \rightarrow \infty$.
%$T=\max(m_A, m_B)$.
\par\noindent \textit{Sketch of proof:} Security is compromised when one of the $M_A$ edge points of Ray's observation pmf is received. The probability of this event is $\frac{M_A}{M_A+M_B-1}$ and as a result:
\begin{eqnarray}
R_s^{(A)}& = & H(\bs{S}_A|\bs{Y}) = \tau-\frac{M_A}{M_A+M_B-1}\rightarrow \tau,\\
R_s^{(B)}& = & H(\bs{S}_B|\bs{Y}) = \tau-\frac{M_A}{M_A+M_B-1}\rightarrow \tau.
\end{eqnarray}

\par\noindent\textit{Theorem:} We assume that Alice and Bob use PAMs of sizes $M_A$ and $M_B$. Setting $T=\max(m_A, m_B)$ and using a triple binning approach we can achieve all secrecy rates in the convex hull delimited by the pairs $R_s^{(A)}$ and $R_s^{(B)}$:
\begin{eqnarray}
&&R_s^{(A)}\leq\tau - 2^{-K_A},\\
&&R_s^{(B)}\leq\tau - 2^{-K_B},\\
\text{with }&& K_A , K_B \in  [0, T-\tau], K_A+K_B\leq T-\tau.
\end{eqnarray}
The proof is omitted due to space limitations. In Fig. \ref{fig:Generic rates} the achievable secrecy rates are depicted for $M_A=2, 4, 8, 16, 32, 64, 128, 256$ and $M_B=64$.

Finally, viewing an $M$-QAM modulator as two orthogonal $\sqrt{M}$-PAM modulators results in the doubling of the secrecy rates when $2$-dimensional modulators are employed, e.g., when Alice employs a QPSK and Bob an $M_B$-QAM modulator ($M_B\geq 4$) then asymptotically $2$ secret bits can be exchanged in each transmission slot as $M_B$ increases.
The achievable secrecy rates when Alice employs an $M_A$-QAM modulator and Bob a $64$-QAM modulator are shown in Fig. \ref{fig:QAM rates}.
\begin{figure*}[t]
\centering
\includegraphics[angle=0,width=0.78\textwidth]{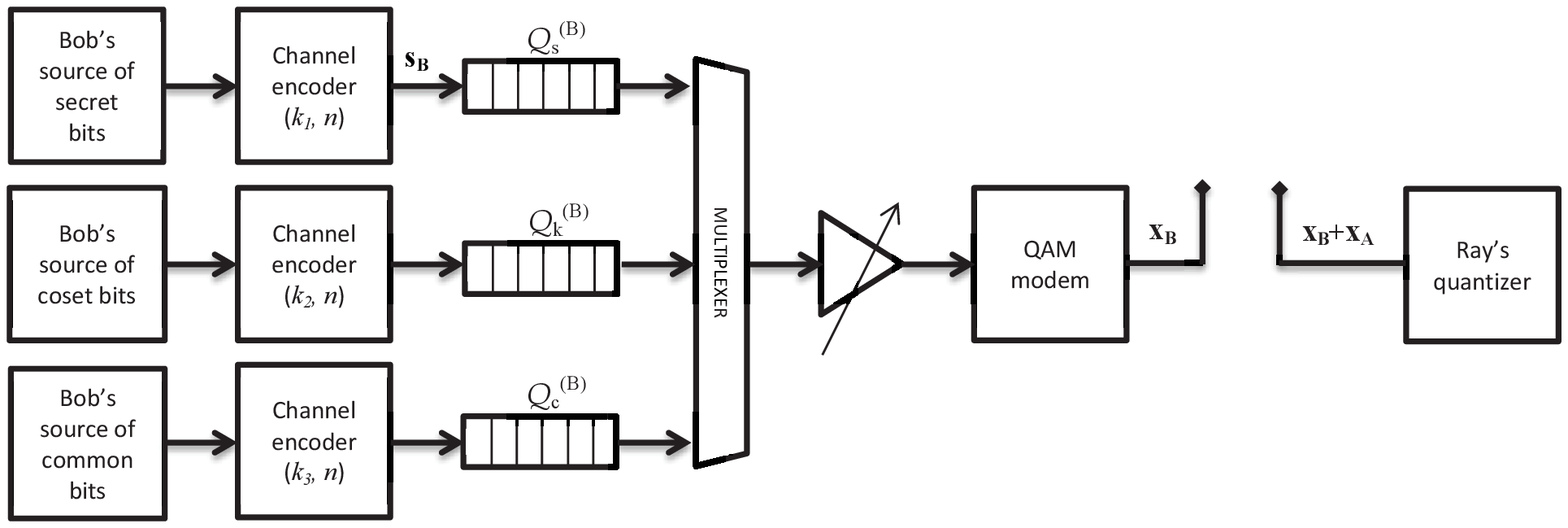}
\caption{Bob's transmitter and Ray's receiver.}
\label{fig:BobTx_RayRx}
\end{figure*}
\begin{figure*}[t]
\centering
\includegraphics[angle=0,width=0.68\textwidth]{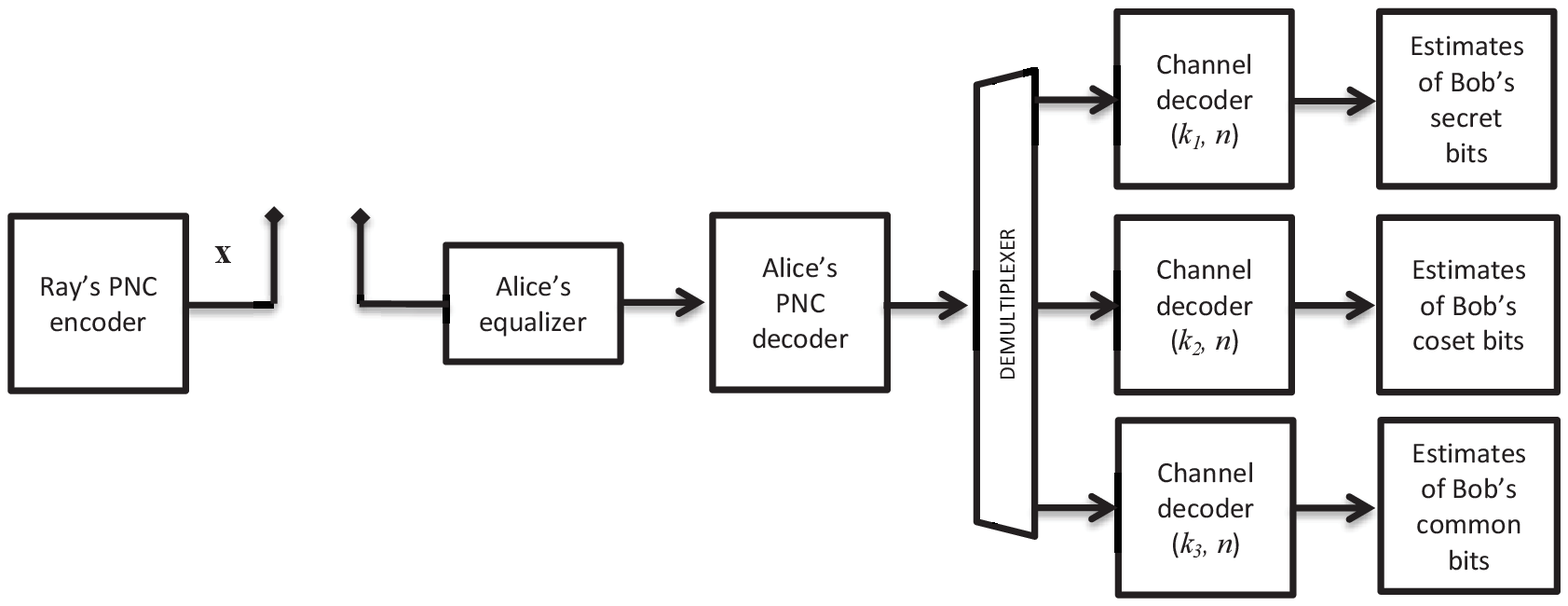}
\caption{Ray's transmitter and Alice's receiver.}
\label{fig:RayTx_AliceRx}
\end{figure*}

\section{System Architecture in Noisy Channels}
In the noiseless case no error control is required and the transmission of secret messages can take place independently from one slot to the next. However, in the presence of noise and other channel impairments such as large scale fading, channel encoders need to be employed to handle the errors introduced by the wireless channel. As a result, to ensure both reliability and secrecy the transmission of secret messages is spread over $n$ subsequent slots as described in section \ref{sec:system model}.

In the schematic diagram in Fig. \ref{fig:BobTx_RayRx} a proposal for Bob's (resp. Alice's) transmitter and Ray's receiver are shown. Three independent block channel encoders are inserted at the outputs of Bob's common, secret and index bits sources to introduce the necessary redundancy to handle the channel errors. Furthermore, as shown in Fig. \ref{fig:RayTx_AliceRx}, at Alice's (resp. Bob's) receiver the corresponding channel decoders are used. In the proposed approach, the tasks of reliability and secrecy are handled independently by separate parts of the transceiver. In the future we will investigate the use of lattice encoders to propose a comprehensive PNC approach that jointly achieves reliability and secrecy.

\section{Conclusions and Future Work}
A scheme that achieves strong secrecy was presented using standard $M$-QAM modulators. We introduced a novel triple binning approach in the largest of the QAM constellations. In our approach, the QAM symbols are generated as the concatenation of i) a bin that carries information bits (secret or common), ii) a bin that carries index bits intended for Bob and iii) a bin that carries index bits for Alice. We have shown that on a per symbol basis it is possible to asymptotically transmit as many secret bits as the length of the shortest of the QAM symbols. Furthermore, accounting for real channels we proposed the use of independent block encoders to alleviate the effects of noise and fading, while the possibility of using lattice encoders will be investigated in the future. Finally, alternative power allocation schemes will also be examined.

\bibliographystyle{IEEEtran}
\bibliography{ICCC14}

\balance

\end{document}